# Systematic study related to the role of initial impurities and irradiation rates in the formation and evolution of complex defects in silicon for detectors in HEP experiments[*]

S. Lazanu[1], I. Lazanu[2]

[1] National Institute for Materials Physics,
POBox MG-11, Bucharest-Magurele, Romania
e-mail: lazanu@alpha1.infim.ro

[2] University of Bucharest, Faculty of Physics,
Bucharest-Magurele, POBox MG-11, Romania
email: i_lazanu@yahoo.co.uk

**Abstract**

**The influence of oxygen and carbon impurities on the concentrations of defects in silicon for detector uses, in complex fields of radiation, characteristic to high energy physics experiments, is investigated in the frame of the quantitative phenomenological model developed previously by the authors and extended in the present paper. Continuous irradiation conditions are considered, simulating realistically the environments for these experiments. The generation rate of primary defects is calculated starting from the projectile - silicon interaction and from the recoil energy redistribution in the lattice. The mechanisms of formation of complex defects are explicitly analysed. Vacancy-interstitial annihilation, interstitial and vacancy migration to sinks, divacancy, vacancy- and interstitial-impurity complex formation and decomposition are considered. Oxygen and carbon impurities present in silicon could monitor the concentration of all stable defects, due to their interaction with vacancies and interstitials. Their role in the mechanisms of formation and decomposition of the following stable defects: $V_2$, $VO$, $V_2O$, $C_i$, $C_iO_i$ $C_iC_s$ and $VP$, is studied. The model predictions cover a generation primary rate of defects between $10^2$ pairs/cm$^3$/s and $10^{11}$ pairs/cm$^3$/s, and could be a useful clue in obtaining harder materials for detectors for space missions, at the new generation of accelerators, as, e.g. LHC, Super-LHC and Eloisatron, or for industrial applications.**

Keywords: silicon, detectors, radiation damage, defect kinetics and impurity dependence, phenomenological model predictions

PACS:   29.40.-n: Radiation detectors
        61.80.Az: Theory and models of radiation effects
        61.82.Fk: Semiconductors
        61.72.Cc: Kinetics of defect formation and annealing

---

[*] Work in the frame of CERN - RD50 Collaboration

## 1. Introduction

Silicon-based devices have found applications in a wide variety of hostile radiation environments. For this reason, radiation effects in silicon represent a research field that have received extensive attention, in order to assess the performances of silicon to radiation. Various systematic studies have been performed to understand better the origins and consequences of radiation damage in silicon detectors. These cover the production of primary defects [1], their annealing mechanisms [2], [3], [4], [5], [6], [7], [8], the correlation with the characteristics of the irradiation particle [9], [10], [11] and with initial material impurities [12], [13]. The modification of material characteristics determines changes in detector parameters. The microscopic phenomena and their consequences at the device level are still poorly understood.

In the last decade a lot of studies have been carried out to investigate the influence of different impurities, especially oxygen and carbon, as possible ways to enhance the radiation hardness of silicon for detectors - see, e.g. references [14] and [15]. These impurities added to the silicon bulk modify the formation of electrically active defects, thus controlling the macroscopic parameters of devices.

The model developed previously by the authors [16], [17], [18] is extended in the present paper to include the reactions of interstitials with vacancy – complexes, as well as those of vacancies with interstitial – complexes. The paper has as main aim the systematic study of the production and evolution of complex defects in silicon, incorporating different quantities of doping impurities, in different irradiation environments.

Until now, there are only few systematic experimental results in the investigation of the modifications of the structural composition of the semiconductor material, and of their effects on the characteristics of devices, after the irradiation in high energy particle fields. Consequence of defect formation and evolution in the material, some macroscopic effects in the device are produced. .The principal macroscopic effects are: bulk material resistivity modification, change of the depletion voltage, increase of the leakage current, change of the electric field distribution in irradiated silicon p-n junction, decrease of the collection efficiency and decrease of the signal to noise ratio [19], [20], [21]. Thus, the correlation established by the model between the initial material parameters, the rate of defect generation in the continuous irradiation regime, the production of defects and their time annealing, could be a useful clue in obtaining harder radiation materials for detectors.

## 2. The mechanisms of production of complex defects – present status

### Vacancy and interstitial defects

Vacancies and interstitials are generated by two mechanisms: thermally and by irradiation, the thermal generation rate being, on basis of thermodynamic arguments, a function of temperature.

The vacancy has been found to have five charge states in the forbidden gap. Two of these: $V^+$ and $V^-$ have been directly identified by electron paramagnetic resonance (EPR) [22] and the characteristics of all vacancies have been obtained in the simple theoretical model which gives the details of the electronic structure [23]. Experimentally, only the positions of the electrical levels of the vacancy for the states (+/0) and (++/+) have been identified. The annealing is the result of long range migration of the vacancy. Vacancy is instable and interacts with other defects. Vacancies which have escaped recombination with interstitials become mobile around 100 K. The exact temperature and the activation energy associated with this mobility depend on the charge state.

The interstitial is considered to be mobile at all temperatures in silicon. A possible explanation of its athermal migration is the Bourgoin mechanism [24]. It is assumed that the defect can exist in more than one charge state and there is sequential trapping and detrapping of charge carriers generated by the ionization which occurs throughout the period of electron irradiation. There is no direct proof of the existence of isolated interstitials in silicon, and their presence has been deduced from theory or from the existence of interstitial impurities, moved from their sites in the lattice to interstitial positions, in concentrations very large compared to the concentration which could result from their direct displacement by irradiation [22].

Nothing is known experimentally about self interstitial annealing processes. In the old Dienes and Damask's papers [2], [3], the authors proposed the formation of diinterstitial $I + I \underset{K_2}{\overset{K_1}{\rightleftarrows}} I_2$, but this defect was not put in evidence experimentally. Interstitials are associated with possible mechanisms of formation of complex defects. Recently, Lindstrom and co-workers [7] reported the experimental identification in IR studies of the effects of possible annealing reaction initiated by silicon self interstitials (see equations. from 8 to 14, 15 and 17 in their up-cited paper).

### Divacancy

In crystalline silicon bombarded with energetic projectiles, the divacancy centre is being studied for about 40 years by numerous authors applying various experimental techniques, e. g., EPR [25], photoconductivity [26], infrared absorption [27], electron-nuclear double resonance (ENDOR) or deep level transient spectroscopy (DLTS) and at the room temperature it is considered as a stable defect. The unperturbed configuration of divacancy could be viewed as two vacant, nearest neighbour, lattice sites. The formation mechanism is the reciprocal trapping of two migrating vacancies: $V + V \to V_2$.

Most of the irradiation studies performed in the 300-600 K range put in evidence that $V_2$ and VO centre are the most important in undoped or moderately doped silicon. Isochronal and isothermal annealing studies have concluded that the divacancy anneals out at 570K [25]. The mechanisms of $V_2$ annealing are not fully understood. It is generally agreed now that annealing of $V_2$ occurs by dissociation and/or annihilation by an impurity or defect with a concentration at least one order of magnitude higher than the $V_2$. It has been found that annealing of the divacancy-related levels, the singly negative $V_2$(0/–), and the doubly negative $V_2$(–/=) charge states at 220–300 °C results in the formation of a new centre, unknown (noted X) with singly negative, $X$(0/–), and doubly negative, $X$(–/=), charge states. The new centre anneals out at 325–350 °C during isochronal treatment for 15 min. Some authors [7], [28], based on IR absorption and DLTS measurements, found that the reaction $V_2 + O_i \to V_2O$ is a possible mechanism for divacancy annealing out and found a possible dependence of its kinetics on impurity concentrations. Pellegrino and co-workers [29], based on DLTS measurements in electron irradiated and Si implanted Si, found that $O_i$ couldn't be the main trap for divacancies, and that the reaction $V_2 + I \to V$ could be of interest only at very high irradiation fluences. The same reaction has been proposed by Davies [4], who couldn't put in evidence such a mechanism by IR absorption.

### $VP$ - centre

The VP centre is another defect that has been carefully investigated from the remote past: by EPR [30], by IR absorption [31], Hall-effect and electrical-conductivity measurements [32]. It is important in n-type silicon with low oxygen content, being a vacancy trap. It anneals out at around 400 K [33].

## $V_iO_j$ - complex defects

In this class of defects, defects with $i, j = 1, 2, 3,...$ are considered. From these centres, the VO is the best known. $V_iO_j$ complexes generally follow the $V_2$ and/or VO annealing at higher temperature [34]. The VO centre (A centre) is the main radiation defect induced in oxygen rich samples. It is one of the first and most thoughtfully studied defects in silicon and is formed when interstitial oxygen captures a vacancy. $V + O_i \rightarrow VO$. This centre has been put in evidence by EPR [30], IR absorption [31], by Hall-effect and electrical-conductivity measurements [32], by DLTS. The VO centre anneals out at around 620 K [35]. Migration of VO and reaction with $O_i$, with the formation of $VO_2$ pairs ($VO + O_i \rightarrow VO_2$) could be one of the annealing out mechanisms of $V_2$ (Londos [36] identified the $VO_2$ centre after irradiation followed by annealing at temperatures higher than 570 K, while Lindstroem [7] identified $VO_2$ in samples irradiated in the temperature range 300 – 600 K). The $VO_2$ centre anneals out at around 720 K, by the possible mechanism: $VO_2 + O_i \rightarrow VO_3$. $VO_2$ and $VO_3$ are the dominant defects identified by IR absorption after electron irradiation at temperatures in the range 600 – 800 K. In carbon - lean silicon, another defect, the oxygen dimmer, has been identified: A proposed formation mechanism is: $I + VO_2 \rightarrow O_{2i}$ [7], for which the reverse reaction: $V + O_{2i} \rightarrow VO_2$ is also possible. In contrast with this situation, in carbon rich samples the oxygen dimmer has not been identified. Other oxygen – vacancy complexes, as $V_2O_2$ have been identified in the same temperature range, by the same group, supposing the reactions: $V + VO_2 \rightarrow V_2O_2$ and $V_2 + O_i \rightarrow V_2O$, the last as an annealing out mechanism for divacancies.

### Complexes formed with carbon

The Watkins replacement mechanism [37] is the main mechanism responsible for interstitial disappearance after irradiation, especially in n-type silicon. $C_i$ has been identified by EPR and DLTS [38]. Interstitial carbon anneals in at 160 – 180 K, and anneals out at 260 -280 K. $C_i$ reacts with impurities to form stable defects: $C_i + X \rightarrow C_iX$ (X could be oxygen, carbon, phosphorous) [39]. So, $C_i$, mobile at room temperature, is captured by a substitutional carbon to form the $C_iC_s$ complex: $C_i + C_s \rightarrow C_iC_s$, that anneals out most probably by dissociation [40]. The disappearance of $C_iC_s$ centres during further irradiation has been found, and has been attributed to mechanisms of the type: $C_iC_s + I \rightarrow CCI$ and $CCI + I \rightarrow CCII$ [4], but not experimental evidence exists.

$C_i$ could be also be captured by an interstitial oxygen, and form the $C_iO_i$ complex: $C_i + O_i \rightarrow C_iO_i$, identified after irradiation at 300 – 600 K by IR and DLTS [4], [7]; $C_iO_i$ could capture another interstitial, to form the $IC_iO_i$ centre [4]. To explain the observation that $C_iO_i$ centres disappear at further irradiation, the most probable hypothesis is by the capture of newly formed primary defects. For carbon rich samples, in the irradiation temperature range 600 – 800 K, the interstitial carbon interacts with vacancy – oxygen centre and forms complexes: $VO_2 + C_i \rightarrow C_iVO_2$ or $C_sO_{2i}$, and $VO_3 + C_i \rightarrow C_iVO_3$ or $C_sO_{3i}$ [7]. The last two defects are not present in samples irradiated at RT and subsequently annealed.

## 3. Modelling of the evolution of defects

The primary incident particle, having kinetic energy in the intermediate up to high energy range, interacts with the semiconductor material. The recoil nuclei resulting from these interactions lose their energy in the lattice. Their energy partition between displacements and ionisation is considered in accord with Lindhard's theory [41, 42] and after this step the concentration of primary defects is calculated. The basic assumption of the present model is that primary defects, vacancies and interstitials, are produced in equal quantities and are uniformly distributed in the material bulk. They are produced by the incoming particle, as a consequence of the subsequent collisions of the primary recoil in the lattice, or thermally. In authors' model, the generation term ($G$) is the sum of two components: $G_R$ accounting for the generation by irradiation, and $G_T$, for thermal generation. The concentration of the primary radiation induced defects per unit fluence (CPD) in silicon has been calculated as the sum of the concentrations of defects resulting from all interaction processes, and all characteristic mechanisms corresponding to each interaction process, using the explicit formula from reference [18]. Due to the important weight of annealing processes, as well as to their very short time scale, CPD is not a direct measurable physical quantity.

In silicon, the primary defects, vacancies and interstitials, are essentially unstable and interact via migration, recombination, and annihilation or produce other defects. In some previous papers (see, e.g. refs. [43], [16], [17] and [18]), the authors developed successively a quantitative phenomenological model to explain the production of primary defects and their time evolution toward stable defects. The model has been applied to silicon with different quantities of initial impurities, considering different rates of production of primary defects and different conditions of irradiation. Without free parameters, the model predicts the absolute values of the concentrations of defects and their time evolution.

In the description of chemical reactions, the formation and evolution of defects in silicon, around room temperature, could be described as follows:

| | |
|---|---|
| $V + I \underset{G}{\overset{K_1}{\rightleftarrows}} 0$     (1a) | $V + V \underset{K_5}{\overset{K_4}{\rightleftarrows}} V_2$     (2a) |
| $I \overset{K_2}{\rightarrow} \text{sinks}$     (1b) | $I + V_2 \overset{K_6}{\rightarrow} V$     (2b) |
| $V \overset{K_3}{\rightarrow} \text{sinks}$     (1c) | |

| | | |
|---|---|---|
| $V + P \underset{K_7}{\overset{K_4}{\rightleftarrows}} VP$     (3a) | $V + O \underset{K_9}{\overset{K_4}{\rightleftarrows}} VO$     (4a) | $I + C_s \overset{K_1}{\rightarrow} C_i$     (5a) |
| $I + VP \overset{K_8}{\rightarrow} P$     (3b) | $I + VO \overset{K_{10}}{\rightarrow} O_i$     (4b) | $V + C_i \overset{K_{11}}{\rightarrow} C_s$     (5b) |

| | | | |
|---|---|---|---|
| $V + VO \underset{K_{12}}{\overset{K_4}{\rightleftarrows}} V_2O$     (6a) | $I + V_2O \overset{K_{13}}{\rightarrow} VO$     (6b) | $C_i + C_s \underset{K_{15}}{\overset{K_{14}}{\rightleftarrows}} C_iC_s$     (6c) | $C_i + O_i \underset{K_{16}}{\overset{K_{14}}{\rightleftarrows}} C_iO_i$     (6d) |

The reaction constants $K_i$ (i = 1, 4 ÷ 16) have the general form: $K_i = C \cdot v \cdot \exp(-E_i/k_BT)$, with $v$ the vibration frequency of the lattice, $E_i$ the associated activation energy and $C$ a numerical constant that accounts for the symmetry of the defect in the lattice. The reaction constants related to the migration of interstitials and vacancies to sinks could be expressed as: $K_j = \alpha_j \cdot v \cdot \lambda^2 \cdot \exp(-E_j/k_BT)$, with j = 2 (interstitials) and 3 (vacancies), $\alpha_j$:- the sink concentration and $\lambda$ - the jump distance.

The reactions are grouped into three categories. The first is related only to primary defects and their reciprocal interactions (equations 1 and 2); the second involves the reactions of primary defects with impurities and the reactions of decomposition of complexes (equations 3, 4 and 5). Into the last group, the interactions between complexes and of complexes with primary defects are comprised - reactions represented by equations 6. In the present analysis, the reactions considered previously in references [16], [17], [18], have been completed with processes 1c, 2b, 3b, 4b, 5b and 6b, so the model becomes symmetric in respect to the processes initiated by vacancies and interstitials. One could remark that for the defects $V_2$, $VP$, $VO$, $V_2O$, two distinct ways of decomposition are considered: the inverse process to that by which these are formed, and a second way, by interaction with an interstitial, contrary to $C_iC_s$ and $C_iO_i$ defects that could be decomposed only by dissociation and to $C_i$, which could interact with a vacancy to come back into a substitutional position (5b). The difficulty in modelling these processes is due to the fact that most of the activation energies associated are not available from the experimental literature. The only one is the activation energy associated with process 4b, the value considered is 1.7 eV, in agreement with the value reported experimentally by Londos' group [44]. In these circumstances, the other 3 activation energies associated with processes 1c, 2b, 3b and 6b are introduced as parameters. The following numerical values have been assigned to the activation energies: $E_6 = E_{10} = E_{11} = 1.7$ eV; $E_8 = 1.5$ eV, $E_{12} = 2$ eV.

Complexes of other impurities with primary defects could also be added to this reaction scheme.

The individual defects separated by one or more lattice constants could make up clusters, and cannot be treated as independent. The cluster should be regarded as one big defect that probably introduces a high number of electronic states into the bandgap. The number of electrons or holes it could trap is limited to quite a few because of carrier repulsion. Until now, there exist more ambiguities in relation to cluster characteristics, their experimental identification and their macroscopic effects. The existence of clusters is not considered in the defects kinetics in the present model.

In the analysis that follows, high resistivity silicon ($10^{14}$ cm$^{-3}$ Phosphorus): which incorporated between $2\times10^{15}$ and $2\times10^{18}$ O/cm$^3$, and between $3\times10^{15}$ and $3\times10^{18}$ C/cm$^3$ is investigated. The generation rates of primary defects of interest are situated in the range $10^2$ VI pairs/cm$^3$/s ÷ $10^{11}$ VI pairs/cm$^3$/s. This range covers different possible classes of applications; the lowest rate corresponds to the radiation field produced by cosmic rays for orbits in the neighbourhood of the Earth, at an altitude of about 380 km, where the dominant particles are the protons [45]. Higher rates (~ $7\times10^8$ VI/cm$^3$/s) estimate the radiation field produced by hadrons in the region of tracker detectors at the LHC accelerator [46], a multiplicative factor of ten is assigned to the expected environment in the planned Super-LHC accelerator [47], and the highest rate, separated by another factor of ten, corresponds to the Eloisatron respectively. For details of the hypothesis in which these rates have been calculated, see reference [46].

A special discussion must be made for process (1c). It has been found [48] that five orders of magnitude separate the diffusion constant of interstitials and vacancies respectively in silicon at room temperature, this being a direct consequence of the fact that the energy of migration of the vacancy is significantly higher than that of interstitials. In these circumstances, the migration of vacancies to sinks could be neglected.

For generation rates up to $10^{11}$ VI/cm$^3$/s, for the considered material impurities and for 20°C, the relative differences in the concentration of defects between the two variants of the model (without and with reactions 2b, 3b, 4b, 5b and 6b) are insignificant. The model is not so sensitive to the values of the activation energies of reactions 2b, 3b and 4b, 5b and 6b: a decrease of 0.3 eV in the activation energies in respect to the values listed above produces changes up to 0.05 % in the values of the concentrations.

## *4. Results, discussions and predictions*

The temperature is an important parameter that influences radiation damage. Because in most of the experimental and theoretical studies the temperature of 20°C is the reference temperature, the present predictions will be done for this case and the results could be compared with the predictions for lower temperatures, recommended for concrete technical solutions.

**4.1 Model predictions and experimental data**

An important problem in all model predictions represents the confidence level between the experimental data and theoretical calculations. Unfortunately, until now, there are only few measurements of defects production and annealing in the continuous irradiation regime, and only for low rates of irradiation. In these circumstances it is very difficult to evaluate the range of applicability of the model prediction. For the available data, the good agreement between our estimations and the measurements at low fluences and low rates of continuous irradiation with 2.5 MeV and 12 MeV electrons, 1 MeV neutrons was clearly demonstrated in a previous paper [46].

**4.2 The time evolution of the concentration of primary defects with respect to the irradiation rate and the material characteristic**

In order to put in evidence the role of oxygen and carbon doping in the radiation hardness of silicon, we will compare the time evolution of the concentrations of defects for samples with four types of doping: low oxygen and low carbon content - $2x10^{15}$ $O/cm^3$, $3x10^{15}$ $C/cm^3$ (nO, C<), high oxygen and low carbon - $2x10^{18}$ $O/cm^3$, $3x10^{15}$ $C/cm^3$ (O, C<), low oxygen and high carbon - $2x10^{15}$ $O/cm^3$, $3x10^{18}$ $C/cm^3$ (nO, C>) and high oxygen and carbon concentrations - $2x10^{18}$ $O/cm^3$, $3x10^{18}$ $C/cm^3$ (O, C>). These materials are considered to be continuously irradiated at 20°C, during ten years, at rates corresponding to near Earth cosmic protons, LHC and Eloisatron (considered having a generation rate ten times higher than the predicted one for S-LHC).

Vacancies and interstitials are generated in equal quantities. Their time evolution is very different for different irradiation rates, and could be seen in Figure 1.

The time evolution of vacancies is controlled, at the lowest generation rate considered here by the rate of generation and by the oxygen content. The concentration of vacancies is higher in non-oxygenated samples, while in oxygen rich material it is lowered due to the formation of VO centres (a decrease by a factor of 250 in the concentration of vacancies for 1000 times higher oxygen). At these irradiation rates, the concentration of interstitials is practically independent both of oxygen and of carbon doping. Because the processes initiated by interstitials are much faster than those corresponding to vacancies, the concentration of interstitials remains practically constant in time and equals the thermal equilibrium value.

In the irradiation conditions characteristic to the LHC, the situation is quite different. Both the decrease of the concentration of vacancies in oxygenated materials and of interstitials in carbon-rich samples is predicted by model calculation. The difference in the concentrations of vacancies due to the content of carbon is low, as is the influence of oxygen on interstitial concentration.

At the highest generation rate considered here, the differences in the concentrations of oxygen and carbon impurities are relevant in the concentrations of vacancies and interstitials only in the first $50 \times 10^6$ seconds of irradiation, at longer times the concentrations of vacancies and interstitials being practically constant. In these conditions, the processes that can be initiated by self-interstitials are favoured due to the higher values of interstitial concentrations. So, the range of higher rates of irradiation where the concentration of interstitials remains high enough could represent more favourable conditions to test the formation of complex defects involving carbon, as, e.g.:
$C_iC_s + I \rightarrow CCI$ and $CCI + I \rightarrow CCII$, if these processes exist.

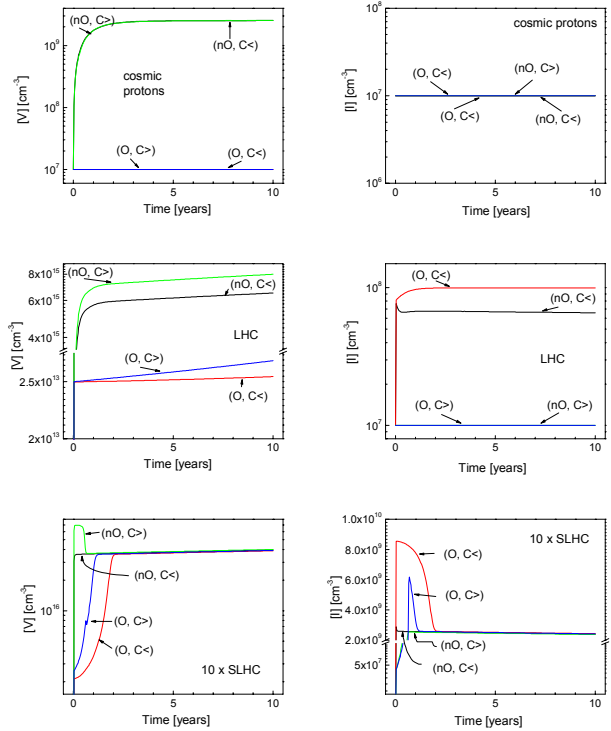

Figure 1
Time dependence of the concentrations of primary vacancies and interstitials, for continuous irradiation with generation rates corresponding to cosmic ray protons ($G_R \sim 10^2$ VI/cm$^3$/s), to hadrons in the central tracker at LHC ($G_R \sim 10^9$ VI/cm$^3$/s), and for an order of magnitude higher rate than the predicted one for S-LHC ($G_R \sim 10^{11}$ VI/cm$^3$/s), for four different silicon compositions (see text).

The regime of high rates is also important to put in evidence the reactions of decomposition of vacancy-impurity defects, initiated by interstitials, of the type 2b, 3b, 4b, 6b.

## 4.3 The correlation between the concentration of stable defects, of impurities and the irradiation rates

The defects considered here are: $V_2$, $VO$, $V_2O$, $C_i$, $C_iO_i$, $C_iC_s$, and $VP$. These could be classified into three categories: divacancies, complexes primary defect - impurity, and only between impurities.

A general feature of the time evolution of the divacancy is the continuous increase of its concentration in time. At long times, a linear increase of the divacancy concentration in time is obtained, the slope being determined by the generation rate of primary defects.

At low rates of generation, divacancy concentration is correlated only with oxygen concentration and is insensitive to carbon. Higher concentration of oxygen conduces to a decrease of the concentration of divacancies with a ratio of about $10^6$ if the oxygen concentration is increased with 3 orders of magnitude. This behaviour is maintained if the rate of generation of primary defects increases with up to seven orders of magnitude. Minor differences in the concentrations of divacancies have been obtained for different concentrations of carbon. At rates of generation of the order of $10^{11} cm^{-3} s^{-1}$, the concentration of divacancies decreases with the increase of oxygen content and with the decrease of carbon, but at long enough times the concentration of divacancies becomes only slightly dependent on the concentrations of oxygen and carbon.

The time evolution of the concentration of divacancies in the three different conditions of irradiation before mentioned, and for four silicon compositions, is presented in Figure 2

.

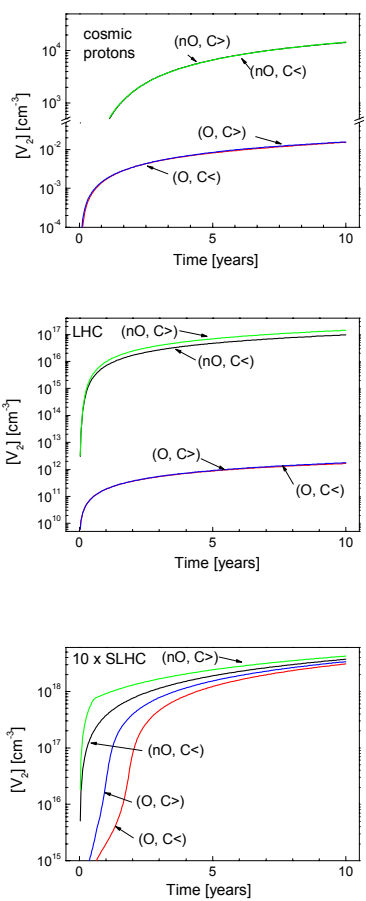

Figure 2
Time evolution of the concentration of divacancies in the same conditions as in Figure 1.

From the impurity – primary defect complexes, we will first discuss the complex related to oxygen - VO, $V_2O$ and $C_iO_i$.

The time dependence of the VO centre presents two different stages: first, there is an accumulation of defects, the processes by which VO forms being preponderant, followed further by a stage where the decomposition is more pronounced.

The time dependencies of both $V_2O$ and $C_iO_i$ concentrations are characterised by a monotonic increase in time and by the possibility to attend a plateau value.

The increase of oxygen content slows down all processes, and also increases the weight of the $C_iO_i$ centre. All processes are faster at higher generation rates.

The relative weight of the concentrations of each of these defect centres in respect to the oxygen concentration in silicon is illustrated in Figure 3, for continuous irradiation at 20°C, in LHC conditions of generation rate, for non-oxygenated ($2x10^{15}$ $O/cm^3$) and oxygenated ($2x10^{17}$ $O/cm^3$) silicon, with $3x10^{17}$ $C/cm^3$.

The maximum value of the ratio [VO]/[O] is around 0.35, independent on the generation rate and on the oxygen concentration in the sample. At low rates (less than $10^8$ $VI/cm^3/s$) and in oxygenated silicon, the ratio [VO]/[O] does not attain its maximum during 10 years, presenting a monotonic increase.

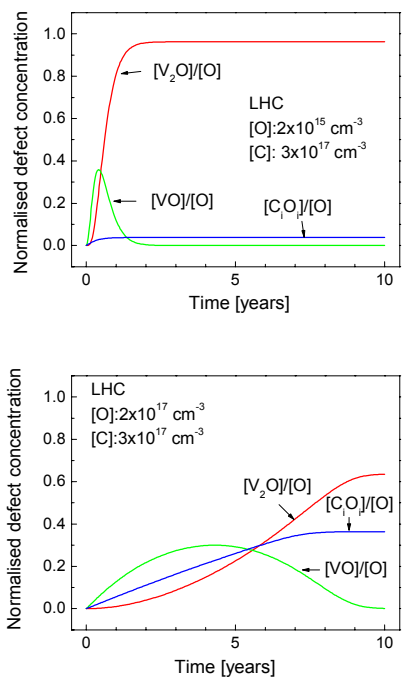

Figure 3
Time dependence of the concentrations of VO, $V_2O$ and CO centres normalised to the concentration of oxygen in silicon, for continuous irradiation with $G_R = 10^9$ VI/cm$^3$/s at 20°C, for non-oxygenated and oxygenated silicon.

At the highest rate of generation considered in this study ($G_R = 10^{11}$ VI/cm$^3$/s), the concentrations of VO, $V_2O$ and $C_iO_i$ attain their equilibrium values, and silicon is stabilised in respect to these defects. This stabilisation corresponds to very low values of the concentrations of the VO centre.

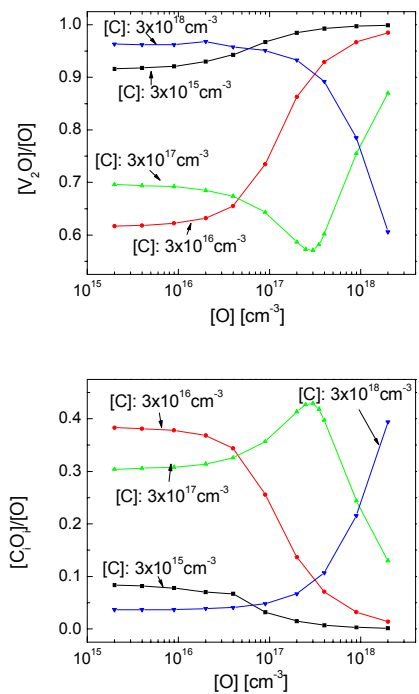

Figure 4
Dependence of normalised equilibrium concentration $[V_2O]/[O]$ and $[C_iO_i]/[O]$ on oxygen concentration for continuous irradiation with $G_R = 10^{11}$ VI/cm$^3$/s.

In Figure 4, the ratio of the stationary concentrations of the $V_2O$ and $C_iO_i$ centres respectively, to oxygen concentration, as a function of the concentration of oxygen is represented. The points represent calculated values, while the lines are drawn only to guide the eye. It can be seen that both oxygen and carbon play important roles. There is a threshold carbon concentration, around $10^{17}$ C/cm$^3$, where the dependencies change.

Carbon associated defects studied here are: $C_i$, $C_iO_i$ and $C_iC_s$. The relative weight of the concentrations of each of these defect centres in respect to the carbon concentration in silicon is illustrated in Figure 5, for $G_R = 10^9$ VI/cm$^3$/s, for carbon-lean (3x10$^{15}$ C/cm$^3$) and carbon-rich (3x10$^{17}$ C/cm$^3$) silicon, with 2x10$^{16}$ O/cm$^3$. While [$C_iC_s$] and [$C_iO_i$] time dependencies are characterised by a monotonic increase, followed eventually by an equilibrium plateau, [$C_i$] could have a maximum in its dependence, correlated with the irradiation rate, the oxygen and carbon content. The relative weight of carbon distribution among these defects depends strongly on the carbon concentration in the sample.

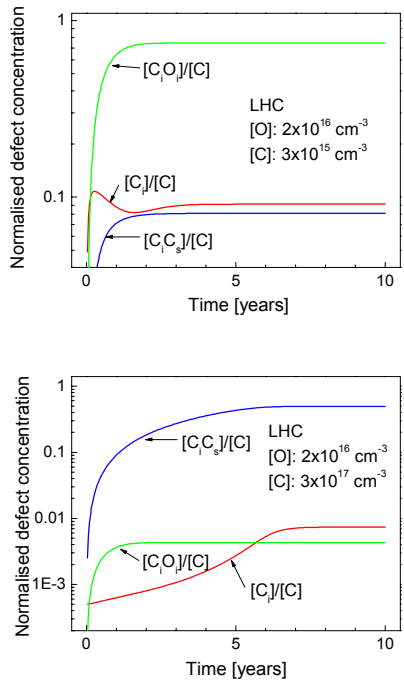

Figure 5
Time dependence of the concentrations of $C_iC_s$, $C_iO_i$ and $C_s$ centres normalised to the concentration of carbon in silicon, for continuous irradiation with $G_R = 10^9$ VI/cm$^3$/s at 20°C, for carbon-lean and carbon-rich silicon.

In this study, the concentration of phosphorus is supposed to be fixed at [P] = $10^{14}$ cm$^{-3}$. In Figure 6, the time dependence of [VP] is represented for continuous irradiation, for two rates of generation: in the conditions of LHC: curves 1, 3, 5 and 7, and for rates one order of magnitude higher than the ones corresponding to S-LHC: curves 2, 4, 6 and 8 respectively. Materials incorporating the concentrations of oxygen of 2 x 10$^{15}$ O/cm$^3$ (curves1, 2 5 and 6), and 2 x 10$^{17}$ O/cm$^3$ respectively (curves 3, 4, 7 and 8) and two concentrations of carbon: 2 x 10$^{15}$ C/cm$^3$ (curves 1, 2, 3 and 4), and 2 x 10$^{17}$ C/cm$^3$ (curves 5, 6, 7 and 8 respectively) have been considered.

After ten years of irradiation, the concentration of the VP centre manifests a tendency to plateau values. These values depend on the irradiation rate, being higher for higher generation rates, and depend slightly on the oxygen content, being lower for oxygenated samples: this is due to the fact that oxygen is a competitor for phosphorous in vacancy trapping. At higher rates, of the order of $10^{11}$ VI/cm$^3$/s, the differences appear only at short times, up to 20 x 10$^6$ s, where the relation between the oxygen and carbon concentration determines the dependence.

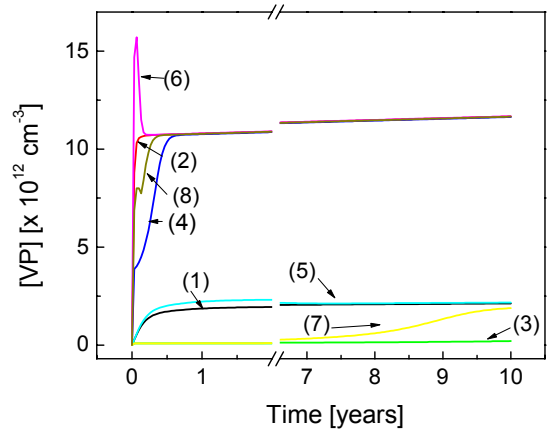

Figure 6
Time dependence of the concentration of VP centres for two rates of irradiation, for non-oxygenated and oxygenated, for carbon – lean and carbon rich silicon – see text for details.

**4.4 Other aspects: consequences of defects production for silicon detectors**

Most semiconductor radiation detectors are based on the properties of the p-n junction. Consequences of irradiation processes, some characteristics of the devices could be influenced by the formation of secondary defects which are unstable and which could be electrically active - with energy levels located in the forbidden band gap of silicon. In this case, they could capture free electrons or holes and change the initial concentrations of charged donors and acceptors in the space charge region of the detector. The principal effects are: change of depletion voltage, increase of leakage current, bulk material resistivity modification, change of electric field distribution in irradiated silicon p-n junction, charge in the collection efficiency and capacitance contribution to the noise.

In this paper, only the modification of the effective concentration of donors and acceptors and some qualitative aspects related to the increasing of the leakage current will be discussed.

The changes of the effective concentration of donors and acceptors ($N_{eff}$) in time during irradiation are responsible for the modification of the depletion voltage. In Figure 7, the time dependence of the effective concentration is represented, for three rates of generation of primary defects, corresponding to cosmic protons exposure, to LHC and 10 times higher rates than Super-LHC, in the conditions of continuous irradiation.

From the defects considered in the model, relevant for modifications of effective concentration are those with deep energy levels in the band gap, in the neighbourhood of the intrinsic level. In accord with reference [20], the deepest level of the divacancy, the level associated with the $C_iO_i$ centre, and the level at $E_c - 0.52$ eV, assigned by us to the $V_2O$ centre have been considered. Shallow donor (P) removal, by the creation of the $VP$ centre has also been considered. The starting material is high resistivity silicon, with $10^{14}$ P/cm$^3$, with different concentrations of oxygen and carbon: (2 x $10^{15}$ O/cm$^3$ and 3 x $10^{15}$ C/cm$^3$), for which the notation (nO, C<) has been used; (2 x $10^{18}$ O/cm$^3$ and 3 x $10^{15}$ C/cm$^3$) – corresponding to the notation (O, C<); (2 x $10^{15}$ O/cm$^3$ and 3 x $10^{18}$ C/cm$^3$) – (nO, C>); and (2 x $10^{18}$ O/cm$^3$ and 3 x $10^{18}$ C/cm$^3$) – (O, C<) respectively. In the model calculations, one could observe a strong correlation between the effective concentration, the inversion phenomenon, the rates of generation of defects and the concentrations of impurities. At lower rates of generation, a linear increase of the effective concentration is evidenced in oxygenated material, the slope depending on the carbon content, while in non-oxygenated silicon this phenomenon is delayed or blocked. In the LHC irradiation conditions, in the oxygenated silicon the inversion is predicted after about 4.5 years of irradiation, but this effect is not present in the non-oxygenated material. At the highest rates of irradiation investigated, the inversion is predicted for all considered materials, faster in those oxygenated. For non-oxygenated silicon, the carbon content monitors the speed of the process. In these circumstances, after one year of irradiation the effective concentration has the tendency to saturate, at values different for different concentrations of impurities.

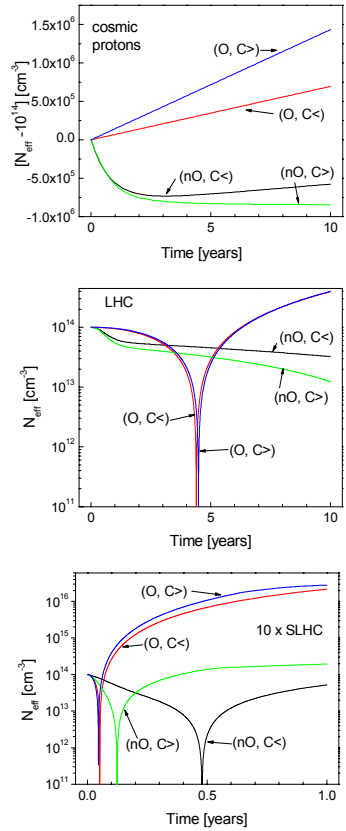

Figure 7
Time evolution of the effective concentration in the space charge region of the p+n junction, for three different generation rates of primary defects due to cosmic rays protons ($G_R \sim 10^2$ VI/cm$^3$/s), to hadrons in the central tracker at LHC ($G_R \sim 10^9$ VI/cm$^3$/s), and for $G_R \sim 10^{11}$ VI/cm$^3$/s, for four differently doped silicon (see the text for compositions and symbols).

Other defect levels, put in evidence by experiment but not assigned to defect centres [20] (and eventually not considered in the model), could be important and could modify the present predictions.

The phenomena of trapping and de-trapping of charge carriers on defect levels are also responsible for the decrease of the charge collection efficiency (CCE): if the charge carrier, produced by the ionising radiation, trapped on a deep level is re-emitted at a time much later than the maximum collection time, it has to be considered as lost.

The increase of the leakage current is due to the generation of electron-hole pairs on the defect levels, and could be evaluated in the frame of the simplified Shockley-Read-Hall model. The defects which have an important contribution to leakage current are those with energy position nearest to the intrinsic level and with the higher cross sections for carrier capture. Because all cross sections are about $10^{-15} \div 10^{-17}$ cm$^{-2}$, the main contributions to current are due to energies levels. So, the most dangerous defects are $V_2$, VP and $V_2$O. Up to generation rates corresponding to LHC, the effect of oxygenation of the silicon is a lower of concentration of detects and consequently one could obtain a decrease of the leakage current in comparison with non-oxygenated material. This advantage is progressively lost for even higher generation rates, so at one order of magnitude higher than the Super-LHC rate, the differences between oxygenated and non-oxygenated silicon are irrelevant. Quantitative predictions of the leakage current for some particular situations have been previously published in reference [18].

## *5. Conclusions*

In the frame of an original phenomenological model developed by the authors, the role of oxygen and carbon impurities in the mechanisms of formation and the kinetics of the following stable defects: $V_2$, $VO$, $V_2O$, $C_iO_i$, $C_i$, $C_iC_s$, and $VP$ in silicon for detector uses, in complex fields of radiation has been investigated supposing continuous irradiation for ten years. The resulting modification of the effective concentration of donors and acceptors in the space charge region of the detector, and some qualitative aspects related to the increase of the leakage current have also been discussed. The obtained results, presented and analysed in the paper, could be suggestions in obtaining harder materials for detectors at the new generation of accelerators, for space missions or for industrial applications.